\documentclass[12pt,preprint]{emulateapj}

\slugcomment{Accepted by the Astrophysical Journal}
\shorttitle{GALEX catalog of AGB stars}
\shortauthors{Montez et al.}
\begin{document}
\title{A Catalog of GALEX Ultraviolet Emission from Asymptotic Giant Branch Stars}
\author{Rodolfo Montez Jr.} 
\affil{Smithsonian Astrophysical Observatory, Cambridge, MA 02138, USA}
\author{Sofia Ramstedt}
\affil{Department of Physics and Astronomy, Uppsala University, Box 516, 75120, Uppsala, Sweden} 
\author{Joel H. Kastner} 
\affil{Chester F. Carlson Center for Imaging Science, School of Physics \& Astronomy, and Laboratory for Multiwavelength Astrophysics, Rochester Institute of Technology, 54 Lomb Memorial Drive, Rochester NY 14623 USA}
\author{Wouter Vlemmings}
\affil{Department of Earth and Space Sciences, Chalmers University of Technology, Onsala Space Observatory, 439 92, Onsala, Sweden} 
\author{Enmanuel Sanchez}
\affil{Department of Physics, University of Cincinnati, 400 Geology/Physics Bldg, Cincinnati, OH 45221 USA}

\begin{abstract}
We have performed a comprehensive study of the UV emission detected from AGB stars by the Galaxy Evolution Explorer (GALEX). 
Of the 468 AGB stars in our sample, 316 were observed by GALEX. 
In the NUV bandpass ($\lambda_{\rm eff} \sim 2310~\AA$), 179 AGB stars were detected and 137 were not detected. 
Only 38 AGB stars were detected in the FUV bandpass ($\lambda_{\rm eff} \sim1528~\AA$). 
We find that NUV emission is correlated with optical to near infrared emission leading to higher detection fractions among the brightest, hence closest, AGB stars. 
Comparing the AGB time-variable visible phased light curves to corresponding GALEX NUV phased light curves we find evidence that for some AGB stars the NUV emission varies in phase with the visible light curves. 
We also find evidence that the NUV emission, and possibly, the FUV emission are anti-correlated with the circumstellar envelope density. 
These results suggest that the origin of the GALEX-detected UV emission is an inherent characteristic of the AGB stars that can most likely be traced to a combination of photospheric and chromospheric emission.
In most cases, UV detections of AGB stars are not likely to be indicative of the presence of binary companions. 
\end{abstract}
\keywords{}

\section{Introduction}

Stars between 0.8 to 8 $M_{\odot}$, including our sun, will go through the asymptotic giant branch (AGB) phase of stellar evolution. 
During the AGB phase, these large ($R\gtrsim 100~R_{\odot}$) and luminous ($\gtrsim10^{3}~L_{\odot}$) stars experience nuclear burning in shells and lose copious amounts of mass at rates reaching up to $10^{-4}~M_{\odot}~{\rm yr}^{-1}$. 
As a result of their cool photospheric temperatures ($T_{\rm eff}<3500$~K) and cool circumstellar envelopes, AGB stars are well-studied in the optical to radio wavelengths. 
Optical emission is primarily used to study pulsations \citep{2017ARep...61...80S,2016ApJS..227....6V}. 
Near- to Mid-infrared emission can be used to probe the dust content, such as the dust composition and mass loss rates \citep{	
1997A&A...324.1059L,2006MNRAS.369..751W}. 
Far-infrared emission begins to probe the circumstellar material which is dominated by dust reprocessing of the stellar photons \citep{2002A&A...382..184M,2012A&A...537A..35C}. 
Bright molecular line emission is present in the sub-mm/mm regime, providing insight into mass loss, envelope expansion, and molecular chemistry \citep{2001A&A...368..969S,2003A&A...411..123G}. 
Longer-wave (e.g., cm through GHz regime) radio emission is often used to study molecular envelope and large-scale magnetic fields through polarized maser emission \citep{2011ApJ...728..149V}. 
In contrast, only weak ultraviolet (UV) emission is expected from AGB stars due to their cool temperatures and dense circumstellar environments. 
As a result, the characteristics of UV emission from AGB stars are poorly studied even though the UV regime may potentially offer an opportunity to study shocks, magnetic activity, and possible binary companions. 

Recently, the detection of samples of AGB stars with the Galaxy Evolution Explorer (GALEX) has revived the subject of UV emission from AGB stars \citep[e.g.,][]{2008ApJ...689.1274S}.
A few AGB stars have been included in spectroscopic studies of luminous cool giants with space telescopes such as the International Ultraviolet Explorer (IUE) and the Hubble Space Telescope (HST) \citep[e.g.,][]{1995ApJ...442..328R,2007AJ....134.1348D}.
Although some studies attribute AGB UV emission to companions \citep[e.g.,][]{2008ApJ...689.1274S}, some UV spectra of AGB stars reveal emission lines that are characteristic of chromosphere radiation, e.g., CII], Mg II, and Fe II.
Overall, past UV spectroscopic observations of cool stars evolving from giants to AGB suggest chromospheric-like radiation persists as stars evolve, possibly reaching a basal-level driven by acoustic waves and/or magnetic activity \citep{1995A&ARv...6..181S,1998ApJ...494..828J,2011MNRAS.414..418P}.
Does the UV emission from AGB stars indicate the presence of such chromospheres or are companions mainly responsible for UV emission? 

In this paper, we present a comprehensive catalog of UV emission from AGB stars as detected by GALEX. 
We used a large sample of AGB stars to assess their UV detection rates. 
We studied spatial distribution of the detected and undetected AGB stars, compared UV fluxes with multiwavelength fluxes (including phased light curves), and considered the few spectroscopic observations of AGB stars acquired by GALEX. 
With this sample of UV observations of AGB stars, we consider the characteristics of UV emission from AGB stars and revisit the question of the origin of AGB star UV radiation. 

\begin{figure}[ht]
\centering \includegraphics[scale=0.55]{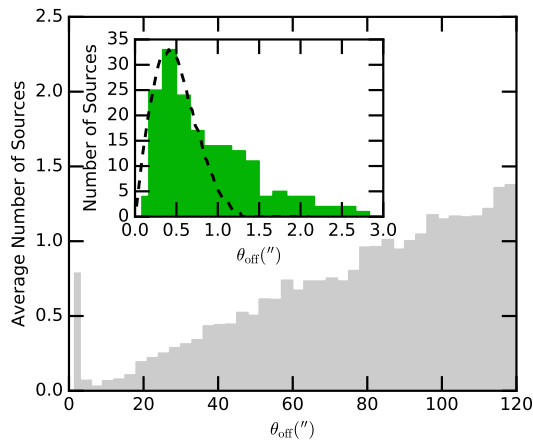}
\caption{Results of AGB/GALEX cross-correlation. Each figure displays the offset, $\theta_{\rm off}$, in arcseconds, of GALEX source positions from the input AGB positions. 
In the main figure, the number of sources are scaled by the number of observations, giving the average number of sources as a function of offset position. 
In the inset, we give the total number of sources (multiple observations included) as a function of offset position and limited to those GALEX sources within $3^{\prime\prime}$ of an AGB star. The dashed line shows the expected distribution based on studies of bright sources. 
\label{detections}}
\end{figure}

\begin{figure*}[ht]
\centering \includegraphics[scale=0.75]{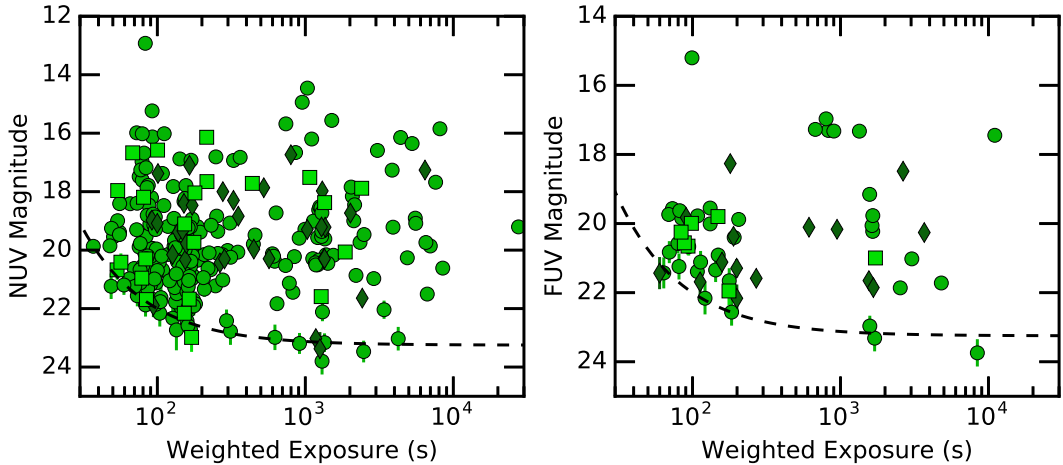}
\caption{NUV ({\it left}) and FUV ({\it right}) observed (reddened) magnitudes as a function of exposure time for all measurements in our sample. In each panel the dashed line indicates the limiting magnitude estimate described the analytic formula $m_{\rm limit}(t_{\rm exp}) = -125~t_{\rm exp}^{-1} + 23.25$~mag and the three chemical sub-types are distinguished as follows: M-type stars as circles, S-type stars as squares, and C-type stars as diamonds. \label{fig:exposures}}
\end{figure*}

\section{Data} 

\subsection{The Sample of AGB Stars}

Our sample of AGB stars is derived from numerous AGB  samples found in the literature and was originally compiled by \citet{2012A&A...543A.147R} to search for X-ray detections associated with AGB stars.
As in \citet{2012A&A...543A.147R}, there are a total of 468 unique AGB stars in our total sample; 286 M-type Miras from \citet{1990AJ.....99.1173L}; 171 AGB stars from the samples of \citet{2001A&A...368..969S}, \citet{2003A&A...411..123G}, and \citet{2006A&A...454L.103R}, plus 11 mixed type stars from the sample of \citet{2008ApJ...689.1274S}. 
The C-type (C/O$>$1) star sample is nearly complete out to 500\,pc. 
The S-type (C/O$\sim$1) star sample is nearly complete out to 600\,pc. 
The completeness of the M-type (C/O$<$1) sample is not well investigated. 

\subsection{GALEX Observations} 

The GALEX mission performed a two-band survey of the UV sky. 
Using dichromatic beam splitter GALEX simultaneously observed FUV ($\lambda_{\rm eff}\sim$1528~\AA;  1344-1786~\AA) and NUV ($\lambda_{\rm eff}\sim$2310~\AA; 1771-2831~\AA) in surveys with different depths. 
The All-Sky Imaging Survey (AIS) had a typical exposure of $\sim$150~s and the Medium Imaging Survey (MIS) had a typical exposure of $\sim$1,500~s. 
The FUV and NUV detectors were photon counting microchannel plates with $\sim1.2^{\circ}$ fields of view with images virtually binned to $1\farcs5$ square pixels.
The spatial resolution is $4\farcs3$ in FUV and $5\farcs3$ in NUV. 
Limiting magnitudes for the AIS are $\sim19.9$ and $\sim20.8$~mag in the FUV and NUV, respectively, for the typical AIS exposure time. 
For MIS, the limiting magnitudes for typical MIS exposure times are $\sim22.7$~mag in the FUV and NUV \citep{2007ApJS..173..682M}. 
GALEX could also perform slitless grism spectroscopy to disperse the FUV and NUV emission. 
As described in further detail in \citet{2007ApJS..173..682M}, spectroscopic observations place a grism into the converging beam of the telescope to simultaneously disperse all sources onto the detector plane. 
According to \citet{2007ApJS..173..682M}, the usable ranges of the grism spectra are 1300-1820~\AA\,  and 1820-3000~\AA\, in the FUV and NUV, with average resolutions of 8~\AA\, and 20~\AA, respectively. 
In May 2009, the FUV detector ceased functioning, but the NUV detector continued functioning well beyond the NASA-led phase, which ended in 2011. 
In all, the GALEX mission made nearly 300 million UV measurements, which are all available via the MAST data archive. 

\subsection{Additional Data}\label{sec:addtnldata}

To supplement our study of the GALEX observations of AGB stars, we collected photometric data from across the electromagnetic spectrum for all the AGB stars considered using SIMBAD and VizieR tools. 
Since many of our stars are bright and exhibit long-period variations, they are often targets of the American Association of Variable Star Observers (AAVSO). 
We collected AAVSO light curves that span the GALEX mission lifetime (05/28/2003 to 06/28/2013) from the AAVSO International Database \citep{aavso}.
Additionally, when available, we have gathered {\it Hipparcos} parallax measurements with signal-to-noise ratios above 1.5 for our entire sample \citep{2007A&A...474..653V}\footnote{Very few stars from our sample were included in Gaia DR1.}. 
To estimate the selective extinction for all bandpasses considered (see \S\ref{detandnondet4p1}) we used the ATLAS9 stellar atmosphere models of \citet{2004astro.ph..5087C}. 

\section{Building the GALEX-AGB Sample}

Positions for the sample of 468 AGB stars were cross-correlated with the GALEX source catalog (General Release 6/7).
Cross-correlation was performed with the Catalog Archive Server Jobs System (CasJobs) using a search radius of $3^\prime$.
This large search radius was used to ensure we included a sufficiently large number of field sources to establish whether an undetected AGB star's field had actually been observed. 
We also limited all sources to those within 0.6 degrees of the center of the field of view. 
The cross-correlation results in a total of 21,603 GALEX sources comprised mostly of field sources with some potential AGB detections. 
Amongst these 21,603 sources 92\% are detected in the NUV, 15\% in the FUV, and only $\sim7\%$ in both NUV and FUV. 
In Figure~\ref{detections}, we display the average number of sources per observation as a function of angular distance from an AGB star, $\theta_{\rm off}$. 
 
The point spread function (PSF) of sources in GALEX NUV and FUV images is measured by the full width at half maximum (FWHM) in two axes and varies as a function of detector position \citep{2007ApJS..173..682M}. 
For high signal-to-noise ratio sources, the ranges of FWHM vary from $\sim5^{\prime\prime}$ near the center of the detector up to $\sim10^{\prime\prime}$ near the detector edges. 
The uncertainty of source positions in the sky has been well-characterized by comparison of bright GALEX sources with Sloan Digital Sky Survey (SDSS) sources \citep{2007ApJS..173..682M}, however, for sources with low signal-to-noise ratios the positional uncertainty is less well-characterized. 
In our cross-correlation results we often find multiple observations and potential counterparts for a given AGB star that are low signal-to-noise GALEX sources. 
The positions of these multiple GALEX detections indicate that for low signal-to-noise sources the positional accuracy can vary up to a few arcseconds. 
In Figure~\ref{detections}, we note a break in the $\theta_{\rm off}$ distribution of GALEX sources at $\sim3^{\prime\prime}$. 
Based on this and on the previous considerations of the PSF and positional uncertainties, we consider any GALEX source within $3^{\prime\prime}$ of an AGB star ($\theta_{\rm off}\leq3^{\prime\prime}$) as a GALEX detection of an AGB star. 
In Table~\ref{tbl:detections} we provide a catalog of all GALEX measurements of AGB stars. Columns 1-6 contain basic information on the AGB star compiled from Simbad including its name, coordinates, chemical type (``Type''), and V and J band magnitudes.  Column 7-14 present information compiled from the GALEX catalog, including an extinction estimate, survey mode, date of observation, the NUV and FUV exposure times and observed magnitudes, and the angular offset between the AGB star position and the GALEX source position. 
The lack of a NUV or FUV exposure and magnitude are indicated by a dash (--) in the respective columns. 

\begin{figure*}[ht]
\centering \includegraphics[scale=0.8]{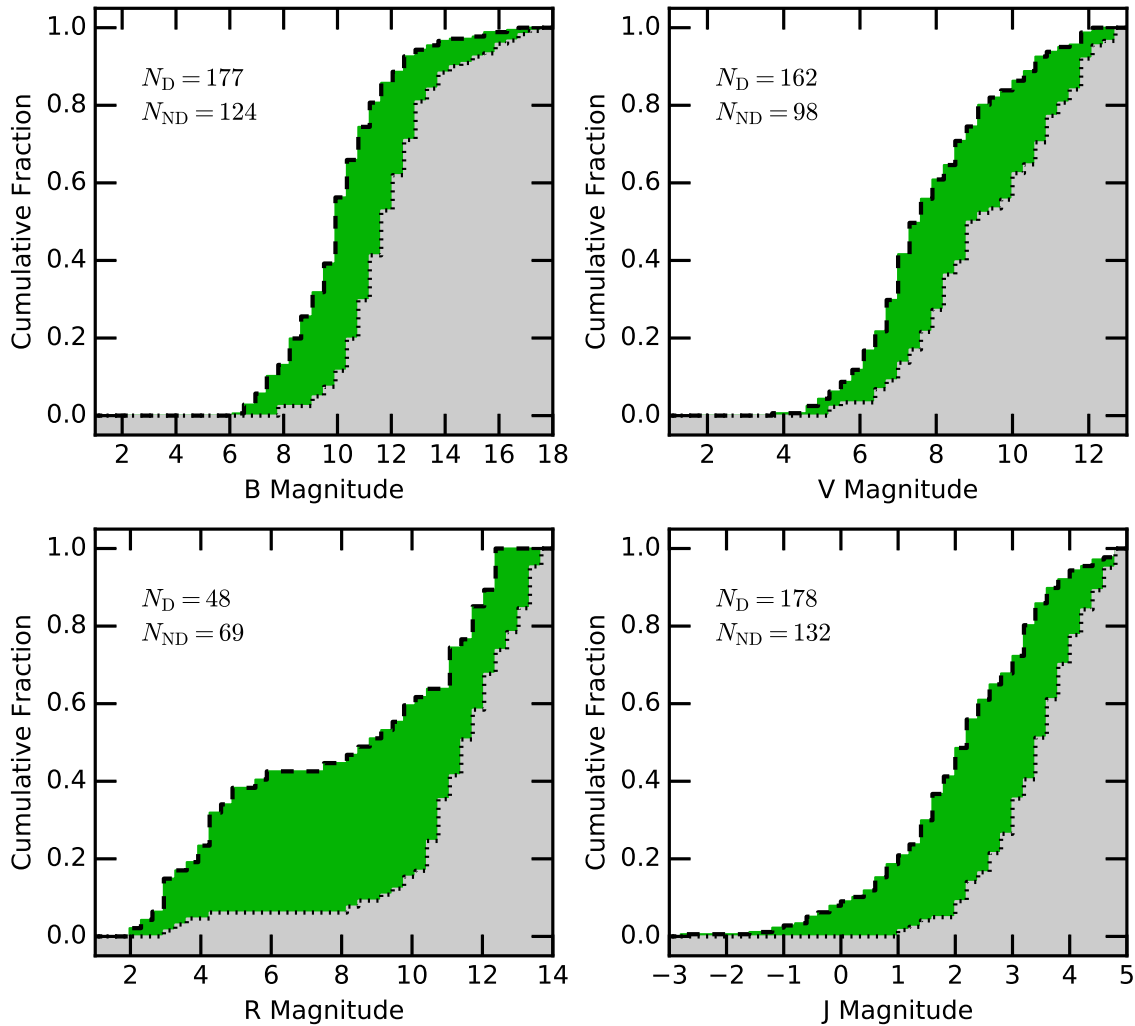}
\caption{Distribution of detections (dashed-line and green-filled) and non-detections (dotted-line and grey-filled). In the four panels we display the cumulative distributions of AGB stars magnitudes in various bandpasses. 
The 2MASS H and K bandpasses (not shown) are similar to the J bandpass. 
\label{fig:brightnessfunc}}
\end{figure*}

To establish AGB non-detections, we based the observation status of a given star on the presence of nearby field sources. 
We considered AGB stars with no field sources within $\sim3^{\prime}$ as {\it unobserved}. 
For an AGB star to be considered as {\it observed} but {\it not detected}, or {\it undetected}, we required that at least one field source lie in the direction of the four major quadrants (NE, SE, NW, SW). 
AGB stars that did not have field sources in all four quadrants were considered {\it uncertain observations} and manually inspected. 
In all 8 cases of the uncertain observations, we determined that the position of the AGB star was off the edge of the detector and thus not observed. 
Table~\ref{tbl:ndstars} lists of all the undetected AGB stars observed by GALEX. 
As in Table~\ref{tbl:detections}, columns 1-6 contain basic information on the AGB star compiled from Simbad including its name, coordinates, chemical type (``Type''), and V and J band magnitudes.  Column 7-11 present information compiled from the GALEX catalog, including an extinction estimate, the NUV and FUV exposure times and observed limiting magnitudes (determined as described in the following paragraphs). 

Given the different depths achieved by the various surveys performed by GALEX, we studied the potential impact of exposure time on detected and undetected sources. 
We plotted the observed magnitudes and effective exposure times of all NUV and FUV measurements of our sample (see Figure~\ref{fig:exposures}). 
These plots show that exposure time varies widely across the sample and suggest that the detectability of the faintest sources is a function of exposure time. 
A majority of the unique detections of AGB stars were made in the shorter exposures that were part of the AIS. 
However, we note that 8 NUV detections of AGB stars and 3 FUV detections of AGB stars are made possible by deeper exposures ($>300$ s); the rest of the deep exposure detections have accompanying detections in the shorter AIS survey exposures. 
Overall, we find that the detections and non-detections of our catalog of AGB stars are unbiased by the various survey depths. 

We used the pattern in Figure~\ref{fig:exposures} to estimate the limiting magnitude as a function of exposure time. 
In particular, we find that a suitable estimate of the limiting magnitude for both the NUV and FUV is described by the simple function $m_{\rm limit}(t_{\rm exp}) = -125~t_{\rm exp}^{-1} + 23.25$~mag. 
For each non-detection, we used the maximum exposure depth in the NUV and FUV bandpasses in the limiting magnitude function to determine the observed NUV and FUV limiting magnitudes (see Table~\ref{tbl:ndstars}). 
Using the same procedure, we estimated limiting magnitudes of non-detections in the FUV bandpass for the stars in Table~\ref{tbl:detections}.

Overall, we find that 316 of the 468 AGB stars in our sample were observed by GALEX.  
Of the 316 observed AGB stars, 179 are detected in the GALEX NUV bandpass and 38 were also detected in the FUV bandpass, while 137 were not detected in either bandpass. 
All 179 AGB stars detected in the GALEX imaging observations, including detections in multiple observations, are listed in Table~\ref{tbl:detections}, and non-detections are listed in Table~\ref{tbl:ndstars}.
In our cross-correlation, we also found 10 AGB stars with GALEX grism spectroscopic observations, which we discuss later. 

\begin{figure*}[ht]
\centering \includegraphics[scale=0.75]{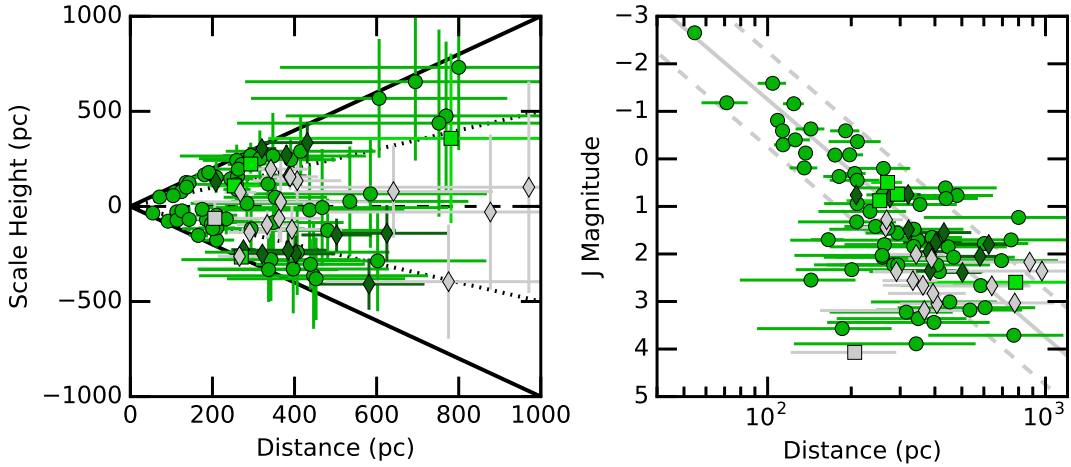}
\caption{{\it Left panel:} Distances and scale heights of detected (green symbols) and undetected (gray symbols) AGB stars with {\it Hipparcos} parallax measurements with SNR$>1.5$. The series of lines indicate galactic latitudes of $b=0^{\circ}$ (dashed), $\pm30^{\circ}$ (dotted), and $\pm90^{\circ}$ (solid). 
{\it Right panel:} J band magnitudes as a function of distance for the same sample. Solid grey line indicates an assumed distance modulus with $(M_{\rm J} + A_{\rm J}) = 6.25{\rm ~mag}$ and $\Delta (M_{\rm J} + A_{\rm J})  = \pm 1.0 {\rm ~mag}$ indicated by the dashed lines. In both panels, the chemical sub-types are indicated as in Figure~\ref{fig:exposures}. \label{fig:scaleheights}}
\end{figure*}

\section{Characteristics of the GALEX-AGB Sample\label{characteristics}} 

\subsection{Detections and Non-Detections}\label{detandnondet4p1}

The characteristics of detections and non-detections amongst the GALEX-AGB sample can provide insight into the nature of GALEX-detected UV emission from AGB stars. 
In Figure~\ref{fig:brightnessfunc} we compare the optical and NIR photometric properties of detected and undetected AGB stars. 
In each panel we display the cumulative distribution of apparent brightness for several optical and infrared photometric bands of the detected and undetected AGB stars\footnote{We only provide the J band of the three 2MASS bandpasses (J, H, K) because they are all very similar in appearance and correlation.}. 
The distributions in Figure~\ref{fig:brightnessfunc} suggest that GALEX-detected AGB stars are approximately two magnitudes brighter, on average, than those that are not detected. 
Given the relatively narrow range of temperatures and bolometric luminosities of AGB stars \citep{1993ApJ...413..641V}, the apparent magnitudes of our AGB stars are a first-order indication of their relative distances. 
Hence the patterns seen in Figure~\ref{fig:brightnessfunc} further suggest that the AGB stars undetected in the UV by GALEX are more distant, on average, than UV-detected stars.
This notion is further supported by the fact that $\sim 44\%$ of the detected AGB stars have {\it Hipparcos} parallax measurements \citep{2007A&A...474..653V} with signal-to-noise ratios $>1.5$, while only $\sim 5\%$ of the undetected AGB stars have similarly significant parallax measurements. 
For the observed J-band apparent magnitudes, which are the least effected by interstellar medium (ISM) extinction of the bands considered, there is clear evidence that the fluxes are proportional to the distance to the AGB stars (Figure~\ref{fig:scaleheights}) and the brighter stars are more easily detected in the UV.  
However, ISM extinction increases towards shorter wavelengths and will influence the overall UV detectability.

\begin{figure*}[ht]
\centering \includegraphics[scale=0.75]{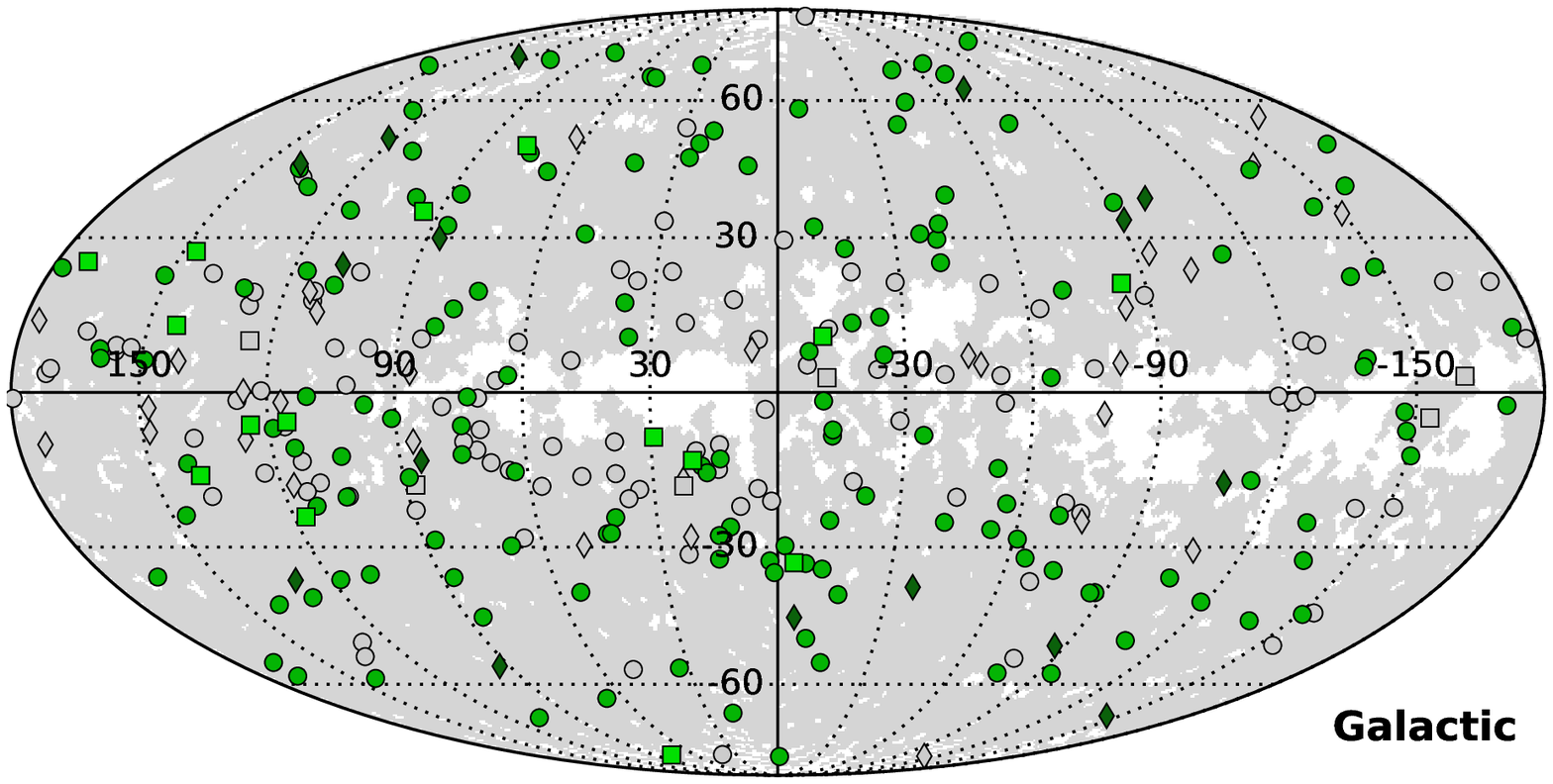} 
\centering \includegraphics[scale=0.75]{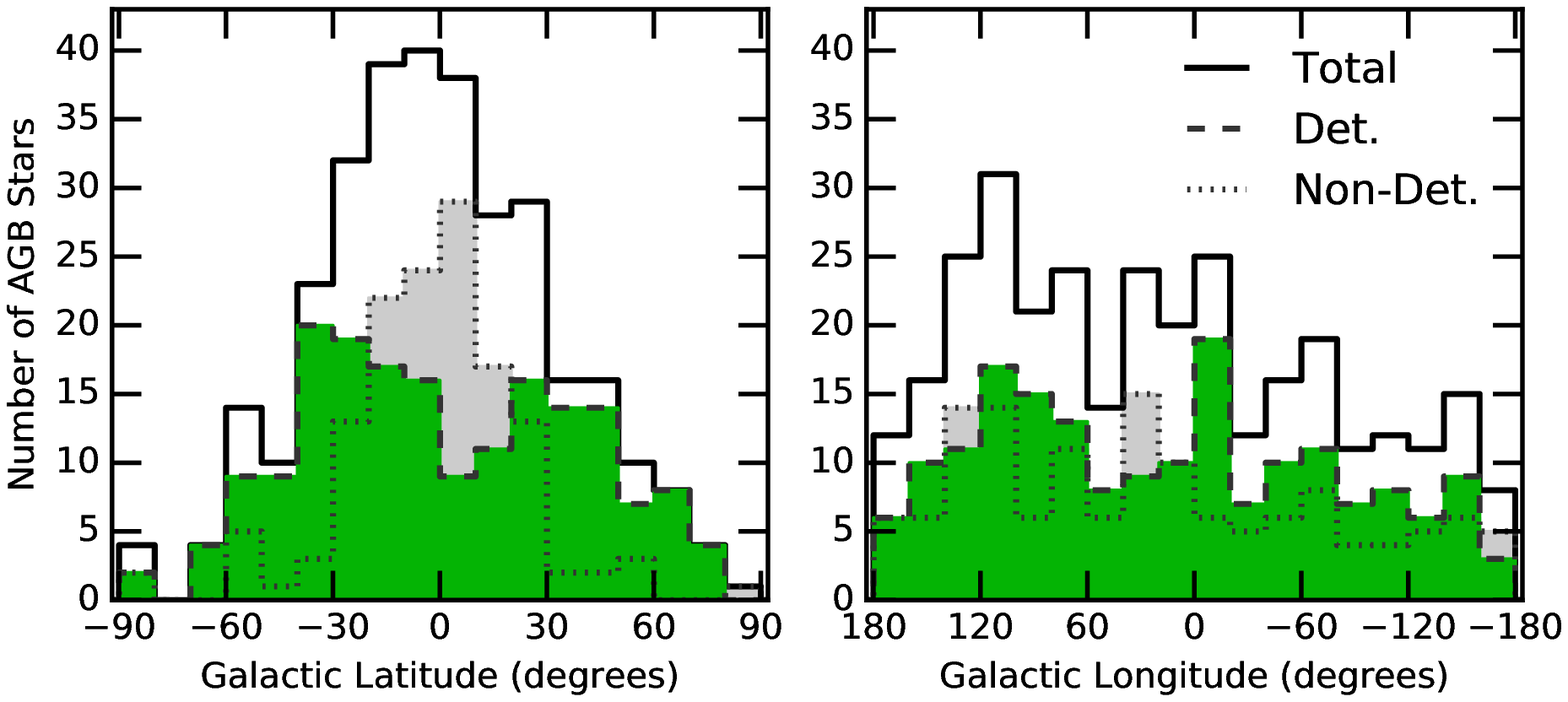} 
\caption{{\it Top panel}: Galactic coordinates ($l$,$b$) of observed AGB stars.  The shaded region represents the approximate coverage for GALEX NUV observations. The chemical sub-types are indicated as in Figure~\ref{fig:exposures}. {\it Lower panels}: distribution of Galactic Latitude ($b$, {\it left}) and Galactic Longitude ($l$, {\it right}) of detected (dashed line filled with green) and undetected (dotted line filled with gray) AGB stars. In both panels the total distributions of the observed AGB stars are indicated by the solid line. 
\label{fig:galdist}}
\end{figure*}

Next, we consider the distribution of the AGB stars (detections and non-detections) in galactic coordinates.
In Figure~\ref{fig:galdist}, we present the distributions in galactic latitude, $b$, and longitude, $l$. 
In galactic latitude, the AGB stars appear to be normally distributed with $\bar{b}\sim0^{\circ}\pm10^{\circ}$ and FWHM$\sim50^{\circ}$.
The non-detections follow a similar distribution but with a narrower FWHM ($\sim25^{\circ}$) compared to the entire sample. 
In contrast, Figure~\ref{fig:galdist} shows that galactic latitudes of the detected AGB stars display a bimodal distribution with a dip in the number of detected stars near $b\sim0$. 
Although GALEX initially avoided low galactic latitudes, more of the galactic plane was observed toward the end of the extended mission \citep{2014Ap&SS.354..103B}.   
Galactic scale heights based on the {\it Hipparcos}-derived distance of detected and undetected AGB stars (see Figure~\ref{fig:scaleheights}) suggest the bimodal distribution of AGB star detections with galactic latitude is not due to the poor coverage of the galactic plane. 
Specifically, we find that AGB stars with high galactic latitudes, $|b|\geq45^{\circ}$, are more readily detected at larger distances than those AGB stars with lower galactic latitudes, $|b|\leq45^{\circ}$. 
Such behavior suggests that high galactic latitudes are more favorable sight-lines for detections. 
This, in turn, suggests that ISM extinction, which increases more rapidly with distance for sight-lines in the galactic plane, is responsible for the decline in AGB star detection fraction with galactic latitude.  
Indeed, that such a trend is due to dust is consistent with studies of larger unbiased samples of GALEX sources such as \citet{2011MNRAS.411.2770B}.  

To mitigate the influence of ISM extinction in subsequent analysis, we use the extinction estimates provided by the GALEX source catalog and listed in Table~\ref{tbl:detections} to deredden the observed magnitudes. 
A caveat of such an approach is that the GALEX extinction estimates are based on galactic dust maps \citep{1998ApJ...500..525S} intended to give total Galactic extinction for a given line of sight through the Milky Way; hence, these extinction estimates are likely to overestimate the extinction for the closest AGB stars that are near the galactic plane. 
Based on the galactic distribution of detections and non-detections, such a reddening correction will be problematic for a majority of undetected sources and $<6\%$ of the detected sources. 
An additional caveat is that the circumstellar material that surrounds an AGB star is still a factor depending on the site of UV emission.  
To estimate the selective extinction in a given bandpass, $A_{\rm BP}/E_{\rm B-V}$, we followed the procedure described in a GALEX study of hot stars \citep{2011Ap&SS.335...51B} for Milky Way-type dust with $R_V = 3.1$.
However, given the nature of our sample,  we used cooler stellar atmosphere models ($T_{\rm eff}<5000$~K) when estimating bandpass selective extinctions. 
For $A_{\rm FUV}/E_{\rm B-V}$ and $A_{\rm NUV}/E_{\rm B-V}$, we determined factors of 7.81 and 6.30, respectively, suggesting that the NUV-FUV color is not independent of extinction for cool stars. 
The selective extinctions for the optical and NIR bandpasses are listed in Table~\ref{tbl:otherbands}.

Next, we compared the detection fraction for the three chemical subtypes (M-, S-, and C-types) indicative of the C/O-ratio in their stellar atmospheres and the dust and molecular chemistry in their circumstellar envelopes\footnote{For M-types, C/O $<$ 1, for C-types, C/O $>$ 1, and for S-types, C/O$\sim$1.}. 
Overall, the observed sample is skewed towards M-types (249), followed by C-types (47) and then S-types (21). 
The detection fractions for our sample of AGB stars are $\sim60\pm5\%$ for M-types, $\sim70\pm20\%$ for S-types, and $\sim34\pm9\%$ for C-types. 
The disparity in M-type versus C-type detection fractions suggests that their different circumstellar environments might influence the UV absorption and, hence, detectability \citep{1992ApJ...398..610M, 2016ApJ...823..104N}.
The lower detection rate amongst C-type AGB stars could be due to their carbonaceous dust, which has higher opacity for photons in the GALEX bandpasses \citep[e.g.,][]{2000MNRAS.315..740S}. 
  
\subsection{Correlation of GALEX-Detected UV Emission with Other Bandpasses}

\begin{figure*}[ht]
\centering \includegraphics[scale=0.8]{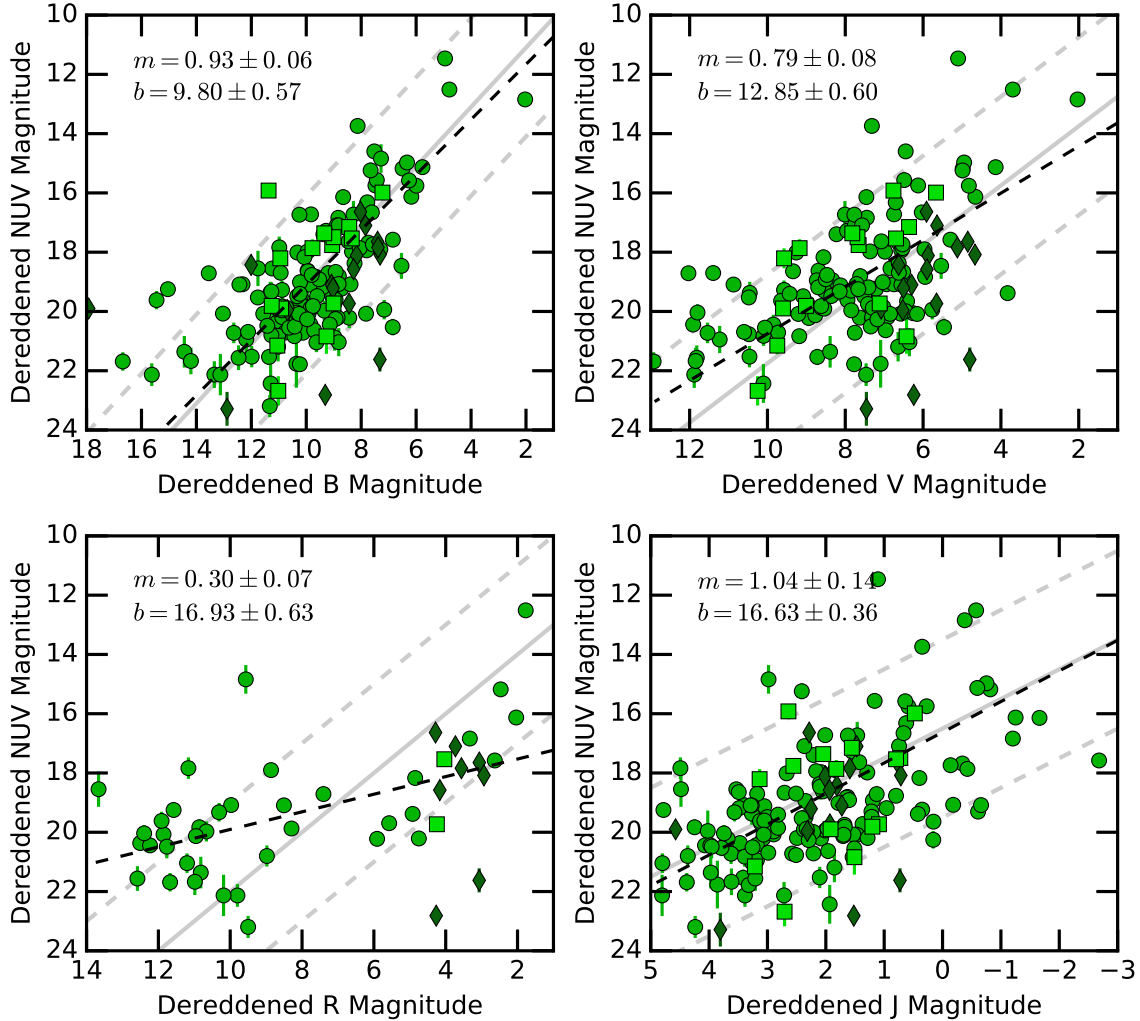}
\caption{Dereddened NUV Fluxes versus B, V, R and J band photometry. 
In each panel, the grey lines represent a simple scaling relationship (slope of unity) with offsets of $-9\pm1$ mag (B-band), $-11\pm3$ mag (V-band), $-12\pm3$ mag (R-band), and $-18\pm3$ mag (J-band). 
The black dashed lines are linear fits of each distribution.
Magnitudes were dereddened using the extinction values in Table~\ref{tbl:detections}. 
In all panels, the chemical sub-types are indicated by distinct symbols as described in Figure~\ref{fig:exposures}.    
\label{fig:nuvfluxes}}
\end{figure*}

\begin{figure*}[ht]
\centering \includegraphics[scale=0.8]{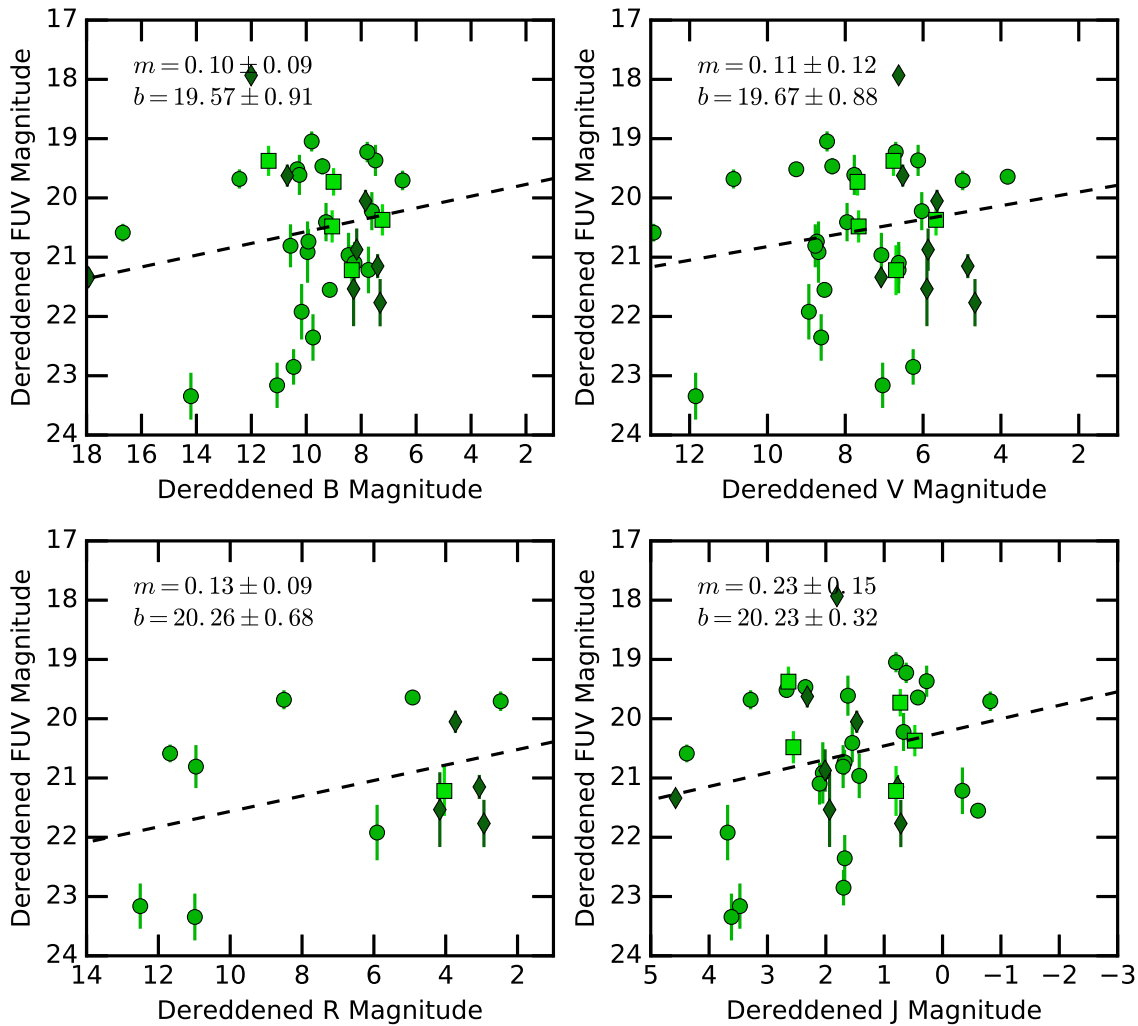}
\caption{Dereddened FUV Fluxes versus B, V, and J band photometry. The four panels display the relationship between the FUV flux with other bandpasses. 
Magnitudes were dereddened using the extinction values in Table~\ref{tbl:detections}. 
The lines and symbols are the same as in Figure~\ref{fig:exposures}. \label{fig:fuvfluxes}}
\end{figure*}

\begin{figure*}
\includegraphics[scale=0.7]{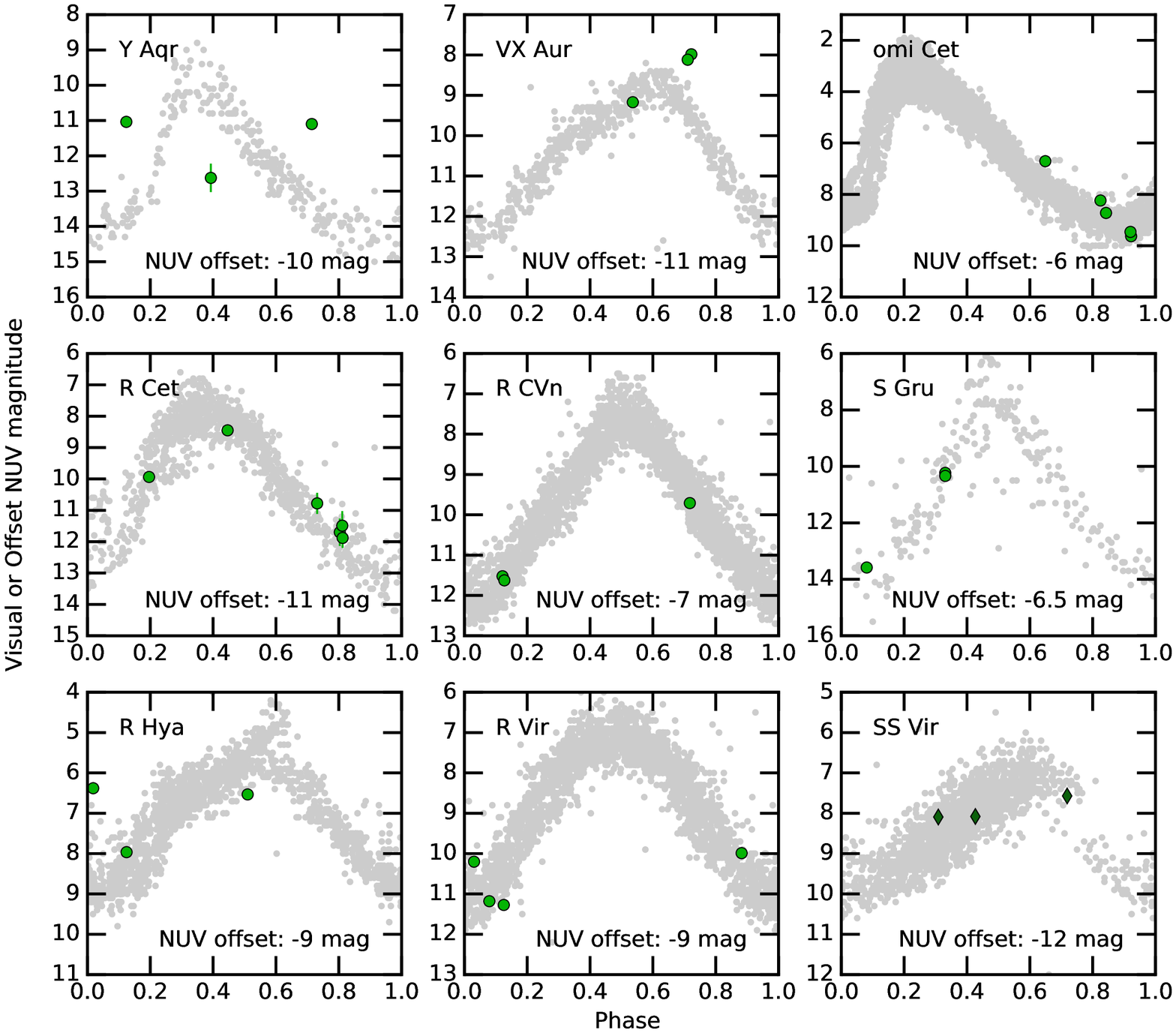}
\caption{A selection of phased AAVSO visual light curves \citep{aavso} of AGB stars with at least three GALEX observations and sufficient AAVSO measurements to clearly display the variability. In each panel we have arbitrarily shifted the dereddened NUV magnitudes (green symbols) to best match the Visual magnitudes (gray symbols). Symbols are the same as in Figure~\ref{fig:exposures}.  \label{fig:lightcurves}}
\end{figure*}

The relationship between the GALEX-detected UV emission and optical and near infrared emission can also help us understand the nature of the UV emission. 
In Figures~\ref{fig:nuvfluxes}-\ref{fig:fuvfluxes} we compare the dereddened optical and near infrared fluxes to the dereddened NUV and FUV fluxes. 
There are apparent correlations between the optical/NIR fluxes and the NUV flux while the FUV flux appears uncorrelated with any of the bandpasses.  
In Table~\ref{tbl:otherbands} we have compiled tests for linear correlation ($r$, the Pearson correlation coefficient\footnote{The Pearson correlation coefficients, $r$, for samples with $N>200$ are robust, but rapidly decrease in significance with smaller sample sizes. }) for the UBVRIJHK broadband photometric bandpasses. 
In all bands considered, we find evidence for correlation with the NUV fluxes and no strong correlation with FUV fluxes. 

Since there is no tight correlation (i.e., no values of $|r| \sim 1$) in the samples shown in Figure~\ref{fig:nuvfluxes}, we considered how the known long-period variability of AGB stars might influence any correlation between NUV and optical/NIR emission. 
In Figure~\ref{fig:lightcurves} we display AAVSO visible light curves \citep{aavso} of AGB stars with at least three separate measurements in the NUV.
In each case, $\sim10$~years of AAVSO measurements acquired during the GALEX mission lifetime were phased to their appropriate periods. 
We phased the dereddened GALEX NUV measurements to the same phase used for the AAVSO light curves. 
Finally, the NUV light curves of a given AGB star were scaled to best match the visible light curve or mean value of the visible light curve. 
As shown in the light curves, the visible light magnitudes for this selection of our sample can vary by up to 8 magnitudes.
  
We note two important properties of the sample of visible and NUV light curves displayed in Figure~\ref{fig:lightcurves}. 
First, the arbitrary scaling used to shift NUV light curves to the visible light curves ranges from 6 to 12 magnitudes. 
Such a large range suggests that multiple scaling relationships exist or that additional influences, such as extinction and possible binary companions, play a role on the level of NUV flux. 
Second, although very few of the examples in Figure~\ref{fig:lightcurves} have adequate phase coverage to absolutely determine the variable nature of the NUV flux, we note that most of the NUV measurements are consistent with the general properties of the visible light curves.  
R Cet has the largest number of observations and largest phase coverage and its NUV light curve clearly mimics the visible light behavior. 
On the other hand, the UV measurements of Y Aqr, which has the next largest phase coverage, appear anti-correlated with the visible light curve. 

Given the range of V-band variability and suspected correlation between GALEX UV emission and the V-band (see Figure~\ref{fig:nuvfluxes}), we attempted to quantify the scatter introduced by non-contemporaneous UV and V-band observations with Monte Carlo simulations of the observations.  
First, we created synthetic V magnitudes drawn from the distribution of V-band magnitudes given in Figure~\ref{fig:nuvfluxes}. 
Next, we used the visual light curves in Figure~\ref{fig:lightcurves} to an estimate the mean amplitude variation of the visible magnitudes ($\Delta {\rm Vis.}\sim4.6$~mag) and their standard deviation (1.6~mag). 
Assuming a similar variability in the NUV (fully-correlated signals), we determine the synthetic NUV magnitudes using a range of power-law scaling relationships characterized by power-law index, $\alpha$, plus a random offset based on the V-band mean amplitude variation. 
Given the well-behaved sinusoidal behavior of the variation seen in Figure~\ref{fig:lightcurves}, we model the variation as a simple uniform random variable in the range of $4.6\pm1.6$~mag. 
With these assumptions, we are able to generate a synthetic sample of NUV and V-band observations that have a 2-D distribution similar to the observed sample seen in Figure~\ref{fig:nuvfluxes}.
We find that a linear scaling relationship ($\alpha=1$) with the given amplitude variation reproduces the sample scatter. 
Scaling relationships with $\alpha<1$ can reproduce the scatter but only if the amplitude variation is increased. 
However, for $\alpha < 0.5$, increasing the amplitude variation fails to reproduce the observed scatter.  
Overall, these considerations suggest that the 2-D distribution of dereddened NUV and V band magnitudes is consistent with correlated fluxes observed non-contemporaneously. 
A more precise determination of the scaling relationship between NUV and V magnitudes requires contemporaneous multiwavelength observing campaigns.  

\begin{figure*}
\centering
\includegraphics[scale=0.7]{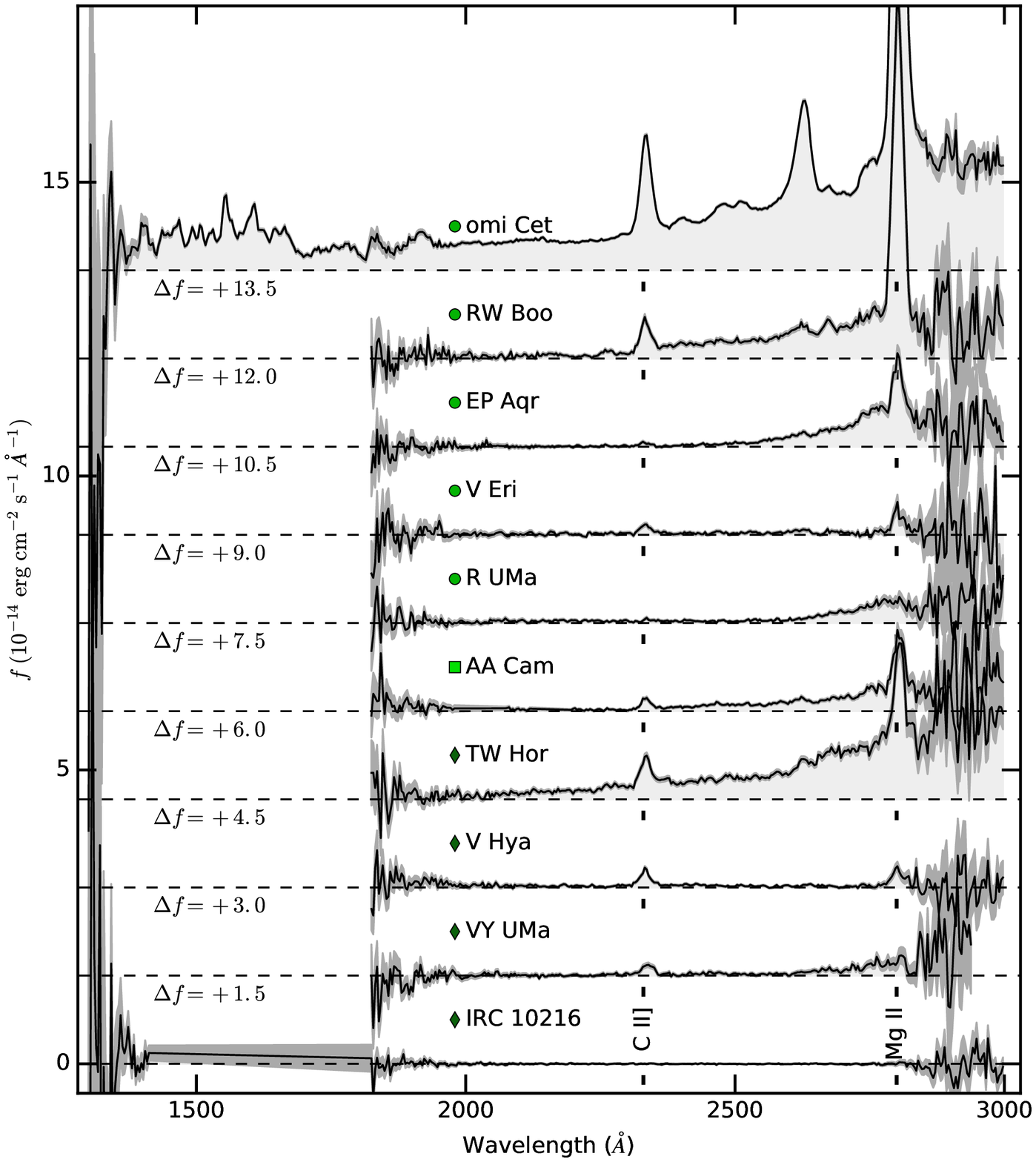}
\caption{GALEX grism spectroscopy of AGB stars. We display the observed grism spectroscopy performed by GALEX for all eight AGB stars with grism observations. 
There is no correction for reddening and spectra are vertically offset by adding the $\Delta f$ quantify indicated in the figure. We indicate $1\sigma$ errors with dark shaded regions and indicate the locations of a few spectral features. Chemical subtype symbols that appear to the left of the object name are the same as in Figure~\ref{fig:exposures}.  \label{fig:spec}}
\end{figure*}

\subsection{Spectroscopy}

Ten AGB stars in our sample were observed with the slitless GALEX grism (see Table~\ref{tbl:grism} for grism observations information). 
All objects with spectroscopic observations were detected in the NUV grism bandpass except for IRC +10216; only Mira is detected in both NUV and FUV grism bandpasses. 
Collectively, the observed spectra are fairly homogenous, displaying some continuum emission with emission lines such as Mg II ($\sim2800$~\AA) and CII] ($\sim2325$~\AA). 
For some sources the continuum may not be detected throughout the NUV grism bandpass leading to a ``flat'' appearance towards longer wavelengths. 
We note that the emission lines are multiplets but the grism spectroscopy has insufficient resolution to resolve the individual lines. 
In Figure~\ref{fig:spec} we display each grism spectrum along with the location of these two common UV emission lines. 
Only two stars, R UMa and VY UMa, do not display the Mg II emission lines, while R UMa also displays a distinct broad feature in the vicinity of the Mg II location. 

\section{Discussion} 

We now consider the possible origin(s) of the UV emission.  
There are two main categories for the possible origin(s) of the UV emission: extrinsic and intrinsic origins. 
Extrinsic origins might include scattering of interstellar radiation field (ISRF) by the circumstellar shell or processes related to hot main sequence or evolved companions to the AGB stars, such as hot photospheres or accretion disks.
Intrinsic origins include the photospheric and chromospheric radiation from the AGB star. 
Although shocks in the circumstellar material are possible, these shocks are not expected to reach UV-emitting temperatures so we do not consider them. 
In the following sections we discuss each of these possible origins given the observed UV detection rates and characteristics of the GALEX-AGB sample. 

\subsection{Extrinsic Origins} 

\subsubsection{Scattering of the Interstellar Radiation Field} 

Scattering of UV photons from the interstellar radiation field (ISRF) by the dusty circumstellar envelope of an AGB star is a potential extrinsic origin for the UV photons detected from an AGB star. 
It has been suggested that the UV emission from the interstellar radiation field (ISRF) influences the chemistry within the circumstellar envelopes of AGB stars \citep{2008A&A...480..431D}. 
The dust-rich environment means that any scattering would preferentially be in the forward direction and some fraction can be scattered into our direction. 
We considered the ``standard'' UV ISRF flux \citep{1978ApJS...36..595D} and estimated the maximum ratio of scattering to bolometric luminosities for the sample of AGB stars with reliable distance estimates~(see \S\ref{sec:addtnldata}).  
Based on simple assumptions (e.g., spherical envelope geometry, standard ISM grain scattering efficiencies), we find that unphysically large scattering radii (1000's of $R_{*}$) with 100\% efficiency are required to account for the measured NUV fluxes. 
For the closest objects, such large scattering radii would produce extended UV sources that are not detected in the GALEX images. 
Also, if scattering of the ISRF is a dominant process for UV emission from AGB stars, then the scattering conditions implied by the NUV suggests that FUV fluxes should be brighter than observed by two orders of magnitude. 
Additionally, it is difficult to reconcile the UV spectral signatures of the grism observations (continuum with emission lines; Figure~\ref{fig:spec}) with scattering of the ISRF. 
We conclude that although scattering of the ISRF is expected to be present, it is unlikely to be a significant source of the UV emission from AGB stars. 

\subsubsection{Binary Companions} 

UV emission from AGB stars has been suggested as a potential tool to detect binary companions from AGB stars \citep{2008ApJ...689.1274S,2016MNRAS.461.3036O}. 
Over most of the electromagnetic spectrum luminous AGB stars will outshine main sequence and/or post-AGB companions. 
However, because the photospheric radiation of an AGB star is expected to drop rapidly towards short wavelengths ($<2800$\AA), hotter companions (main sequence or post-AGB) can dominate short wavelength emission, especially in the NUV and FUV bandpasses. 
\citet{2008ApJ...689.1274S} used such an argument to target a sample of $\sim25$ bright AGB stars, most of which had the ``multiplicity'' flag in the {\it Hipparcos} astrometric catalog.
Amongst this sample, which the authors acknowledge was predisposed towards suggesting the presence of companions, UV excesses were detected amongst 21 of the 25 AGB stars considered \citep[12 were detected in NUV only, and 9 in both the NUV and FUV;][]{2008ApJ...689.1274S}. 
\citet{2008ApJ...689.1274S} concluded that the excesses could not be explained by photospheric radiation and proposed two possible binary origins: photospheric radiation from a companion and accretion onto a companion star. 
Independent of our study, \citet{2016MNRAS.461.3036O} studied a volume-limited ($<500$ pc) sample of 53 AGB stars detected by GALEX. 
This study includes the samples of \citet{2008ApJ...689.1274S} and ``confirmed'' binary AGB stars derived from a radial velocity (RV) study by \citet{2009A&A...498..627F}\footnote{\citet{2008ApJ...689.1274S} and \citet{2009A&A...498..627F} remark on the complications that pulsations of AGB stars pose for RV measurements of such stars. Nevertheless, \citet{2009A&A...498..627F} conclude that some of the Mira and semi-regular variables they studied are suspected binaries, based on their RV measurements.}. 
\citet{2016MNRAS.461.3036O} estimated the NUV and FUV flux in excess of photospheric emission based on stellar spectral templates and proposed that if an AGB star had a NUV excess $\geq20$ and/or if an AGB star is detected in FUV, then its UV emission indicates a binary companion.
Applying the UV-based criteria, \citet{2016MNRAS.461.3036O} inferred a binary fraction of $\sim60\%$ for their sample of AGB stars. 
However, for a given putative companion, \citet{2016MNRAS.461.3036O} also found discrepancies among the effective temperatures obtained from their three temperature estimation methods (i.e., NUV excess, FUV excess, and the ratio of these excesses). 
These discrepancies led the authors to conclude that some of the putative companions might be influenced by the AGB star circumstellar material or that the UV emission is produced by an alternative process.

\subsection{Intrinsic Origins} 

\begin{figure*}[ht]
\centering \includegraphics[scale=0.8]{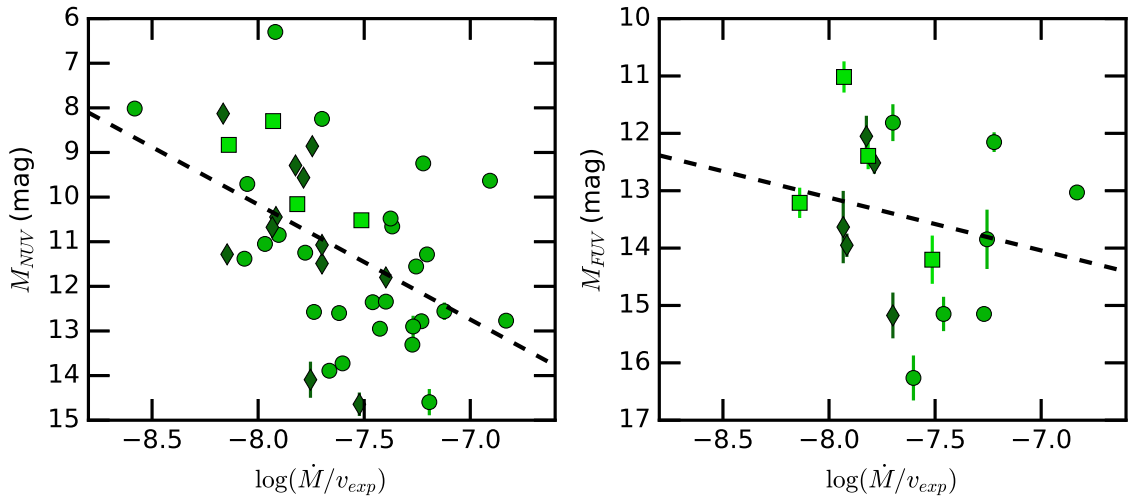}
\caption{Mass loss density proxy and GALEX distance-scaled and dereddened magnitudes for the {\it Hipparcos} sample of AGB stars. The ratio of $\dot{M}$ and $v_{\rm exp}$ is proportional to the density of the circumstellar envelope fed by mass loss. The data indicate that the NUV emission, and to a lesser confidence, the FUV emission are anti-correlated with the circumstellar envelope density. The black dashed lines are linear fits of each distribution.
Magnitudes were dereddened using the extinction values in Table~\ref{tbl:detections}. 
In both panels, the chemical sub-types are indicated by distinct symbols as described in Figure~\ref{fig:exposures}.
\label{fig:mdot}}
\end{figure*}

By analyzing a nearly complete, volume-limited sample of AGB stars, and thereby avoiding the biases inherent in previous studies that employed smaller samples, we find that the brightest (closest and most well-studied) AGB stars have higher detection fractions in the NUV and FUV than more distant AGB stars (\S\ref{detandnondet4p1}). 
This suggests that the GALEX-detected UV emission from AGB stars is most likely inherent to the AGB star and not due to extrinsic processes.
Indeed, IUE observations reveal the prevalence of chromospheric-like UV emission from cool giants and supergiants \citep{2000ApJ...536..923L}. 
The photospheric models adopted by \citet{2008ApJ...689.1274S} and \citet{2016MNRAS.461.3036O} do not include radiation from such overlying chromospheric emission, and it would appear likely such a process is responsible for AGB UV emission in many, if not most, cases. 

The faint UV fluxes, apparent scaling relationships with other bandpasses, and phase correlations between visual and UV magnitudes suggest that the observed UV emission could arise from the Wein tail of the photospheric spectral energy distribution. 
However, we note that the behavior of the photosphere in these regimes is poorly understood and the gas and dust of the circumstellar envelope are strong sources of UV absorption. 
Hence, although photospheric radiation may contribute to the GALEX-detected UV emission, it is unlikely to account for all of the observed UV flux.

Near to far UV spectroscopy from IUE and HST of AGB stars reveals that emission lines from Mg II, CII], and Fe II are prevalent. 
Such emission lines are indicative of a chromospheric layer in the atmospheres of these stars that persists into these highly-evolved stages perhaps never dropping below a basal level \citep{1995A&ARv...6..181S}. 
The chromospheres appear to be much cooler than solar-like chromospheres. 
The heating mechanism is uncertain, but could be tied to pulsations, wave-heating, or magnetic fields \citep[see review by][and references therein]{1995A&ARv...6..181S}. 
The apparent prevalence of UV emission from AGB stars (Figure~\ref{fig:brightnessfunc}), the apparent correlation with stellar fluxes (Figure~\ref{fig:nuvfluxes}), and the spectral characteristics of past IUE and recent GALEX grism spectroscopic observations (Figure~\ref{fig:spec}) suggest that chromospheres are significant contributors to the GALEX-detected UV emission from AGB stars. 

Since the photospheres and chromospheres reside within the circumstellar envelope, the properties of the UV emission should be influenced by absorption by the circumstellar material. 
To explore this notion we compiled mass loss rates, $\dot{M}$, and gas kinematics, $v_{\rm exp}$, for the {\it Hipparcos} sample of AGB stars \citep{1999A&AS..140..197G,2001A&A...368..969S,2003A&A...411..123G,2002A&A...391.1053O,2009A&A...499..515R,2015A&A...581A..60D} to compute the density proxy, $\dot{M}/v_{\rm exp}$.  
In Figure~\ref{fig:mdot} we compare the dereddened and distance-corrected NUV and FUV emission with this mass loss proxy for density. 
Despite the scatter, this comparison provides tentative evidence that the  circumstellar density and the dereddened distance-corrected NUV and FUV flux are anti-correlated. 
Performing a similar comparison for other broad band measurements, we find that the anti-correlation increases for shorter wavelengths. 
Such behavior is expected if the circumstellar material resides between us and the source of UV emission and attenuates the flux. 
The evidence for multiple scaling laws suggested by the NUV correlations with optical and NIR emission could then be explained by varying degrees of absorption in the time-varying circumstellar shell. 

\section{Conclusions} 

We have searched for GALEX detections of a large number of AGB stars culled from various catalogs.  
Among our sample of 468 AGB stars we find that 316 were observed by GALEX, 179 were detected in NUV, and only 38 were detected in both FUV and NUV. 
Based on our analysis of the GALEX data for this sample, we determined the following.
\begin{enumerate} 
\item Comparing the detected and non-detected samples we find that the brightest AGB stars have high NUV detection rates ($>90\%$) and the NUV detection rate decreases with decreasing brightness. 
This suggests that the NUV is an inherent property of all AGB stars and that the brightest and, hence, closest AGB stars are more readily detectable by GALEX.  
Studying the distribution of our sample in galactic coordinates, we find that the non-detections are concentrated toward low galactic latitudes, while the detections exhibit a bimodal distribution in galactic latitude. 
Although GALEX did not completely cover low galactic latitudes, the bimodal behavior seen in the distribution of galactic latitudes is likely due to the increased intervening ISM extinction towards the galactic plane making detection more difficult.  
\item We find that the dereddened NUV fluxes appear correlated with dereddened broad band multi-wavelength photometry (BVRJHK), while the few FUV-detected stars show no such correlation. 
The scatter in these correlations is consistent with that expected from non-contemporaneous observing of correlated time-variable behavior. 
This notion is further supported by analysis of phased AAVSO visual light curves and multiple GALEX observations that suggest the UV emission is correlated with the long-term pulsations observed from these AGB stars.
Overall, these correlations suggest that the emission in the NUV bandpass is inherent to the AGB star.  
Future contemporaneous optical and UV photometric and spectroscopic measurements with good phase coverage are necessary to improve our understanding of the physical mechanism(s) responsible for the UV emission from AGB stars. 
\item UV spectroscopic detections (IUE, HST, and GALEX) of emission lines from AGB stars suggests that chromospheres are present and contribute to the GALEX-detected UV emission from most AGB stars. 
The processes responsible for heating these chromospheres is an unresolved question, but if we assume that the prevalence of UV emission indicates the ubiquity of such chromospheres, then their role in heating and ionizing the circumstellar environment and potentially driving mass loss remain unexplored avenues of future study. 
\item The anti-correlation between the circumstellar density and NUV fluxes further support the notion that the UV emission originates from below the circumstellar shell, consistent with chromospheric and/or photospheric origin of UV from AGB stars. The multiple scaling relationships suggested by the correlations of stellar fluxes with NUV flux can be interpreted as due to various degrees of extinction in the time-varying circumstellar environment. 
\end{enumerate}
The objects in our GALEX-AGB star catalog represent excellent targets for follow-up with UV spectroscopy and contemporaneous optical observations. 
Future simultaneous multi-wavelength observations are urged to verify and expand our understanding of this emerging window into the AGB evolutionary phase. 
Such observations can be used to confirm the origin of the UV emission from AGB stars, to establish the mechanism for chromospheric heating, and to determine the level of magnetic activity at these evolved stages of stellar evolution. 

\acknowledgments
This research has made use of the SIMBAD database, VizieR catalogue access tool, and VizieR photometry tool\footnote{http://vizier.u-strasbg.fr/vizier/sed/} operated at CDS, Strasbourg, France. The original descriptions of the SIMBAD database and VizieR service were published in \citet{2000A&AS..143....9W} and \citet{2000A&AS..143...23O}, respectively. 
We acknowledge with thanks the variable star observations from the AAVSO International Database contributed by observers worldwide and used in this research.
This research made use of Astropy, a community-developed core Python package for Astronomy \citep{2013A&A...558A..33A}.
Some/all of the data presented in this paper were obtained from the Mikulski Archive for Space Telescopes (MAST). STScI is operated by the Association of Universities for Research in Astronomy, Inc., under NASA contract NAS5-26555. Support for MAST for non-HST data is provided by the NASA Office of Space Science via grant NNX13AC07G and by other grants and contracts. 
We are grateful to the anonymous referee for insightful comments and suggestions that improved the manuscript.
WV acknowledges support from ERC consolidator grant 614264.

{\it Facilities:} \facility{GALEX}, \facility{AAVSO}.

\clearpage
\begin{deluxetable}{lrrcccccccccccccc}
\tabletypesize{\scriptsize}
\tablecaption{Catalog of AGB Stars Associated with GALEX Sources \label{tbl:detections}}
\tablewidth{0pt}
\tablehead{
\colhead{Object} & \colhead{RA} & \colhead{DEC} & \colhead{Type} & \colhead{V} & \colhead{J} & \colhead{E(B-V)} & \colhead{Survey} & \colhead{Date} & \colhead{$t_{\rm NUV}$} & \colhead{NUV} & \colhead{$t_{\rm FUV}$} & \colhead{FUV} & \colhead{$\theta_{\rm offset}$} \\
 &  (deg) & (deg) & & (mag) & (mag) & (mag) &  &  & (s) & (mag) & (s) & (mag) & ($^{\prime\prime}$)
 }
\startdata
BC And & 345.22121 & 46.51042 & M & 9.11 & 2.47 & 0.29 & AIS & 2006-11-02 & 132 & $19.23\pm0.08$ & 182 & $>22.6$ & 0.45 \\
BU And & 350.91625 & 39.72692 & M & 10.50 & 2.05 & 0.13 & AIS & 2006-08-07 & 167 & $22.71\pm0.47$ & 192 & $>22.6$ & 2.15 \\
 & & & & & & & MIS & 2011-11-06 & 1297 & $23.80\pm0.45$ & -- & -- & 1.54 \\
R And & 6.00812 & 38.57704 & S & 7.39 & 1.17 & 0.09 & AIS & 2006-11-06 & 83 & $20.30\pm0.16$ & 107 & $>22.1$ & 0.31 \\
SV And & 1.08363 & 40.10995 & M & 7.70 & 3.57 & 0.10 & AIS & 2004-08-19 & 445 & $19.80\pm0.05$ & 178 & $>22.5$ & 0.85 \\
TU And & 8.09471 & 26.02943 & M & 10.22 & 3.42 & 0.04 & AIS & 2003-10-07 & 90 & $21.64\pm0.33$ & 92 & $>21.9$ & 0.53 \\
 & & & & & & & GII & 2004-10-02 & 1211 & $20.50\pm0.06$ & 1610 & $>23.2$ & 0.83 \\
UX And & 38.37001 & 45.65438 & M & 8.69 & 1.31 & 0.11 & AIS & 2006-12-14 & 149 & $20.16\pm0.11$ & 87 & $>21.8$ & 0.76 \\
W And & 34.38734 & 44.30494 & S & 6.70 & 1.59 & 0.09 & AIS & 2006-12-14 & 78 & $20.96\pm0.44$ & 41 & $>20.2$ & 1.12 \\
 & & & & & & & AIS & 2006-12-14 & 84 & $21.69\pm0.37$ & 104 & $>22.1$ & 1.36 \\
 & & & & & & & MIS & 2011-11-23 & 1280 & $21.59\pm0.14$ & -- & -- & 1.57 \\
\enddata
\tablecomments{Table 1 is published in its entirety in the electronic 
edition of the {\it Astrophysical Journal}.  A portion is shown here 
for guidance regarding its form and content.}
\end{deluxetable}

\begin{deluxetable}{lrrcccccccccccccc}
\tabletypesize{\scriptsize}
\tablecaption{Limiting magnitudes for undetected AGB stars with GALEX observations. \label{tbl:ndstars}}
\tablewidth{0pt}
\tablehead{
\colhead{Object} & \colhead{RA} & \colhead{DEC} & \colhead{Type} & \colhead{V} & \colhead{J} & \colhead{E(B-V)} & \colhead{$t_{\rm NUV}$} & \colhead{NUV} & \colhead{$t_{\rm FUV}$} & \colhead{FUV}  \\
 &  (deg) & (deg) & & (mag) & (mag) & (mag) & (s) & (mag) & (s) & (mag)
 }
\startdata
IRC +10011 & 16.60827 & 12.59807 & M & -- & 7.44 & 0.03 & 95 & $>$21.9 & 119 & $>$22.2 \\
IRC +10216 & 146.98919 & 13.27877 & C & 10.96 & 7.28 & 0.05 & 6367 & $>$23.2 & 8422 & $>$23.2 \\
NSV 24833 & 294.75308 & -16.86569 & S & -- & 4.34 & 0.16 & 132 & $>$22.3 & 195 & $>$22.6 \\
LEE 338 & 326.11993 & 73.63468 & C & 9.82 & 3.44 & 0.58 & 380 & $>$22.9 & 60 & $>$21.2 \\
C* 59 & 18.43567 & 62.96006 & C & 9.00 & 4.74 & 1.85 & 74 & $>$21.5 & -- & -- \\
AH And & 31.47750 & 40.72408 & M & 9.30 & 4.16 & 0.07 & 70 & $>$21.5 & -- & -- \\
EY And & 356.25967 & 43.92394 & M & -- & 3.55 & 0.10 & 90 & $>$21.9 & -- & -- \\
KU And & 1.72058 & 43.08333 & M & -- & 3.04 & 0.09 & 72 & $>$21.5 & -- & -- \\
RS And & 358.84059 & 48.63826 & M & 8.38 & 1.54 & 0.15 & 128 & $>$22.3 & -- & -- \\
RY And & 350.15663 & 39.62056 & M & 10.00 & 3.48 & 0.13 & 1366 & $>$23.2 & 198 & $>$22.6 \\
\enddata
\tablecomments{Table 2 is published in its entirety in the electronic
edition of the {\it Astrophysical Journal}.  A portion is shown here
for guidance regarding its form and content.}
\end{deluxetable}

\begin{deluxetable}{lccccccc}
\tabletypesize{\scriptsize}
\tablecaption{Correlations with Other Bandpasses \label{tbl:otherbands}}
\tablewidth{0pt}
\tablehead{
\colhead{Band} & \colhead{$\frac{A_{\rm BP}}{\rm E(B-V)}$} & \colhead{$N_{\rm NUV}$} & \colhead{$r_{\rm NUV}$\tablenotemark{a}} & \colhead{$\rho_{\rm NUV}$\tablenotemark{b}} &  \colhead{$N_{\rm FUV}$}  & \colhead{$r_{\rm FUV}$\tablenotemark{a}} & \colhead{$\rho_{\rm FUV}$\tablenotemark{b}} \\
 }
\startdata
U & 4.85 & 38 & 0.70 & 0\% & 12 & 0.24 & 46\% \\
B & 3.92 &  177 & 0.78 & 0\% & 37 & 0.18 & 28\% \\
V & 3.03 & 162 & 0.63 & 0\% & 38 & 0.16 & 33\% \\
R & 2.26 & 48 & 0.52 & 0\% & 13 & 0.40 & 18\% \\
I & 1.58 & 24 & 0.55 & 0\% & 9 & 0.17 & 65\% \\
J & 0.89 & 178 & 0.48 & 0\% & 37 & 0.25 & 14\% \\
H & 0.56 & 178 & 0.40 & 0\% & 37 & 0.28 & 10\% \\
K & 0.36 & 178 & 0.35 & 0\% & 37 & 0.28 & 9\% \\
\enddata
\tablenotetext{a}{Pearson correlation coefficient for NUV or FUV sample versus given bandpass. In samples with more than 100 values a Pearson correlation coefficient, $r$, above 0.2 is significant evidence of correlation. In samples with fewer than 100 values, the significance of $r$ decreases rapidly. For example, for sample sizes of 20 and 50, $r$ values of 0.3 and 0.45 are not significantly distinct from 0 (no correlation) and $\rho$ becomes unreliable.}
\tablenotetext{b}{Probability that an uncorrelated sample can produce the given distribution. Note: $\rho$ becomes unreliable for small sample sizes.}
\end{deluxetable}

\begin{deluxetable}{lcccl}
\tabletypesize{\scriptsize}
\tablecaption{AGB Stars with GALEX Grism Spectroscopy \label{tbl:grism}}
\tablewidth{0pt}
\tablehead{
\colhead{Object} & \colhead{Obs Date} & \colhead{$t_{\rm NUV}$ (s)} & \colhead{$t_{\rm FUV}$ (s)} & \colhead{Detected} \\
 }
\startdata
EP Aqr & 2007-02-07 & 2759 & 2759 & NUV only \\ 
RW Boo & 2006-11-01 & 5113 & 1721 & NUV only \\ 
AA Cam & 2005-01-30 & 3224 & 1693 & NUV only \\ 
Mira (omi Cet) & 2006-11-18 & 11325 & 11325 & NUV \& FUV \\ 
V Eri & 2005-11-03 & 1704 & 1616 & NUV only \\ 
TW Hor & 2006-05-07 & 1088 & 1007 & NUV only \\ 
V Hya & 2005-08-17 & 2696 & 2696 & NUV only \\ 
IRC +10216 (CW Leo) & 2008-01-23 & 8761 & 8760 & Non-Detection \\ 
R UMa & 2006-01-06 & 1703 & 1703 & NUV only \\ 
VY UMa & 2006-01-07 & 1704 & 1704 & NUV only \\ 
\enddata
\end{deluxetable}

\end{document}